\documentclass[pra,reprint]{revtex4-1}

\usepackage[latin1]{inputenc}
\usepackage{amsmath}
\usepackage{graphicx}
\usepackage{nicefrac}
\usepackage{bbold}
\newcommand{\ket}[1]{\left| #1 \right>} % for Dirac bras
\newcommand{\bra}[1]{\left< #1 \right|} % for Dirac kets
\newcommand{\avg}[1]{\left< #1 \right>} % for average

\begin{document}

% Use the \preprint command to place your local institutional report
% number in the upper righthand corner of the title page in preprint mode.
% Multiple \preprint commands are allowed.
% Use the 'preprintnumbers' class option to override journal defaults
% to display numbers if necessary
%\preprint{}

%Title of paper

\title{Conditioned quantum motion of an atom in a continuously monitored 1D lattice}

\author{Ralf Blattmann and Klaus M{\o}lmer}

%\email[]{Your e-mail address}
%\homepage[]{Your web page}
%\thanks{}
%\altaffiliation{}
\affiliation{Department of Physics and Astronomy, Aarhus University, DK-8000 Aarhus C, Denmark}

%Collaboration name if desired (requires use of superscriptaddress
%option in \documentclass). \noaffiliation is required (may also be
%used with the \author command).
%\collaboration can be followed by \email, \homepage, \thanks as well.
%\collaboration{}
%\noaffiliation

\date{\today}

\begin{abstract}
We consider a quantum particle on a one dimensional lattice subject to weak local
measurements and study its stochastic dynamics conditioned on the measurement outcomes.
Depending on the measurement strength our analysis of the quantum trajectories reveals
dynamical regimes ranging from quasi-coherent wave packet oscillations to a Zeno-type dynamics. We analyse how these dynamical regimes are directly reflected in the spectral properties of the noisy measurement records.
\end{abstract}

% insert suggested PACS numbers in braces on next line
\pacs{03.65.Ta, 03.65.Xp, 42.50.Dv, 42.50.Lc}
% insert suggested keywords - APS authors don't need to do this
%\keywords{}

%\maketitle must follow title, authors, abstract, \pacs, and \keywords
\maketitle

% body of paper here - Use proper section commands
% References should be done using the \cite, \ref, and \label commands

\section{Introduction}

In quantum theory the measurement of an observable leads to
a change of the state of the measured system that depends on the
random measurement outcome \cite{vonNeumann}.
The quantum theory of measurements can include loss and errors and yields a realistic description
of actual measurement processes carried out in a laboratory, including modelling of both projective measurements and of weak (non-projective), continuous measurements \cite{WisemanMilburn}. By continuous measurements, we refer to probing which is not described as an operation acting at a single instant of time, but as the continuous monitoring, e.g., of an optical field  emitted by a quantum system over a finite period of time. The noisy signal from such a measurement is accompanied by a stochastically evolving quantum state of the system, a so-called quantum trajectory \cite{Carmichael}.

While the nature of this measurement back action has been intensively discussed since the beginning of quantum theory \cite{Wheeler}, its consistency with experiments has been verified under different measurement scenarios and in a variety of physical systems \cite{Haroche13,Murch13}. Measurement back-action is, indeed, an efficient way to prepare and control quantum states for which other strategies may not be available \cite{Zeilinger,Polzik,Kuzmich,Nielsen06,Grangier}.

In this article, we study a simple 1D lattice system subject to weak continuous probing
sketched e.g. in the inset of Fig.~\ref{fig:1}(a).  
The system may be implemented as a single particle which is allowed to tunnel among nearest neighbour potential wells in a finite optical lattice or tweezer trap array, and it may also be implemented with a finite chain of spin 1/2 particles with nearest neighbour Heisenberg interactions, prepared with one particle in the spin up state and all the others in their spin down state. We study here the interplay between the evolution of the particle or spin up excitation which becomes delocalized over the lattice due to the tunneling or spin-spin interaction, and the weak probing of a single or a few sites in the model.
For atoms, such probing can be done with a far-off resonant light beam, which experiences a phase shift or polarization rotation depending on the presence or the spin state of an atom. The measurement is weak in the sense that for a very short probing interval, the field contains only few photons, and hence the phase resolution $\Delta \phi \sim 1/\sqrt{n}$ by the correspondingly noisy homodyne measurement does not resolve the atomic states. Integration of the signal over longer times provides better resolution which is, however, in competition with the natural dynamics of the system.

In Sec.~\ref{sec:model}, we introduce the Hamiltonian of the
system and we briefly describe the stochastic master equation that
models the measurement process.
In Sec.~\ref{sec:Results}, we analyse the temporal dynamics of the
system, and we show how frequency analyses of the noisy measurement signal for different probing strengths reveal different regimes for the interplay between the free evolution and the measurement back-action. Finally, in Sec.~\ref{sec:Summary} we summarize
our findings and discuss possible generalizations.

\section{The model}
\label{sec:model}
\subsection{The system}

We consider a single quantum particle on a one dimensional chain with $N$ sites,
described by the tight-binding Hamiltonian
\begin{equation}
  \label{eq:Ham}
  H = J\sum_{n=1}^{N-1} \ket{n+1}\bra{n} + \ket{n}\bra{n+1},
\end{equation}
where $\{\ket{n}\}$ denotes the single site basis (either for the location of a single particle tunneling among the sites, or for a spin up excitation, exchanging location by interacting with neighbouring spin down particles).
In our model we assume degeneracy of the energy of the localized particle or spin excitation over all sites, and since the number of (spin-up) particles is conserved, the dynamics only depends on the coupling parameter $J$. For convenience, we shall describe our results with the terminology for a single particle tunneling between sites, but the results
apply equally to the spin chain or other equivalent systems.

The Hamiltonian Eq.~\eqref{eq:Ham} can readily be diagonalized, yielding the eigen-energies % Reference???
\begin{equation}
  \label{eq:eval}
   e_k = 2J\cos\frac{\pi k}{N+1} % maybe put this inline
\end{equation}
and the corresponding eigenstates
\begin{equation}
  \label{eq:evec}
   \ket{w_k} = \sqrt{\frac{2}{N+1}} \sum_{n=1}^N\sin\frac{\pi k\ n}{N+1}\ket{n},
    % maybe put this inline
\end{equation}
with $k=1,\dots,N$.
Note, that for for odd $k$ the eigenstates
have even symmetry with respect to the middle of the chain,
while for even $k$ they are antisymmetric (odd).

\subsection{Conditioned dynamics}

We consider the situation where the single site population
of the lattice is continuously probed by a coherent light beam
interacting dispersively with the atom. The resulting phase
shift is monitored within a standard homodyne detection scheme.
In order to simulate the dynamics induced by the continuous weak probing of
a Hermitian system observable $O$, we employ the theory of continuous measurements
which, conditioned on the random measurement outcome,
\begin{equation}
  \label{eq:record}
  \lambda_t[O] = \avg{O}_\rho dt + \frac{dW_t}{\sqrt{8k}}
\end{equation}
describes the time evolution of the system density matrix $\rho(t)$ by the stochastic master equation \cite{WisemanMilburn}
% COMMENT: use stochastic Schroedinger equation?
%
%
%\begin{equation}
% \label{eq:SSE}
% d \ket{\Psi} = \Bigl[- \frac{i}{\hbar} H  dt - k\bigl(X - \avg{X}\bigr)^2 dt + \sqrt{2k} \bigl(X - \avg{X}\bigr) dW  \Bigr] \ket{\Psi}
%\end{equation}
%
%
\begin{align}
  \label{eq:SME}
  d\rho = -&\frac{i}{\hbar} [H,\rho]dt \notag \\ +& k \mathcal{D}[O] \rho\ dt
  + 4 k\mu \mathcal{H}[O]\rho\ (\lambda_t[O]-\avg{O}_\rho dt),
\end{align}
where $dW_t$ is a Wiener noise increment with zero mean and variance $\text{Var}(dW_t) = dt$
and $\avg{\dots}_{\rho}=\text{tr}[\dots\rho(t)]$ denotes the average with respect
to the density matrix $\rho(t)$.
In Eq.~\eqref{eq:SME} the Lindblad term
\begin{equation}
       \label{eq:Lindblad}
       \mathcal{D}[O]\rho = 2 O \rho O^{\dagger} - \{O^{\dagger}O,\rho \}
\end{equation}
accounts for a deterministic decoherence of the system due to the measurement, and
the stochastic term with
\begin{equation}
       \label{eq:back}
       \mathcal{H}[O]\rho = O \rho + \rho O - \avg{O+O^{\dagger}}_{\rho} \rho
\end{equation}
represents the information gain associated with the measurement outcome  \cite{WisemanMilburn}.

The parameter $k$ in Eq.~\eqref{eq:SME} accounts for the measurement strength
and $\mu$ is the detector efficiency. In this work, we will restrict ourselves
to the situation where $\mu=1$. In this case, notwithstanding the appearance
of the dissipative Lindblad term in Eq.~\eqref{eq:SME}, the stochastic master equation
for an initially pure states is equivalent to a stochastic Schr{\"o}dinger
equation and preserves the purity of the state $\rho$.

In this work, we consider the situation of local, non-destructive probing of the presence of the particle at a given site, e.g., by a dispersive interaction with an optical field. If the  $n$th-site is monitored, the measured observable is $O = \Pi_n=\ket{n}\bra{n}$. 
% define "outline"?%A stochastic master equation of the form $\eqref{eq:SME}$
%COMMENT: Write sth. about homodyne detection ???
%

\section{Results}
\label{sec:Results}

In this section, we present results of simulations of the evolution of the system subject to weak probing of a single site in the lattice. We assume that, initially, the system is prepared in the eigenstate with the lowest energy, i.e.~$\rho(t=0)=\ket{w_1}\bra{w_1}$.
Hence, the dynamics we observe is excited by the continuous, local probing of the system.

\begin{figure}
\raggedright
  \includegraphics[scale=1]{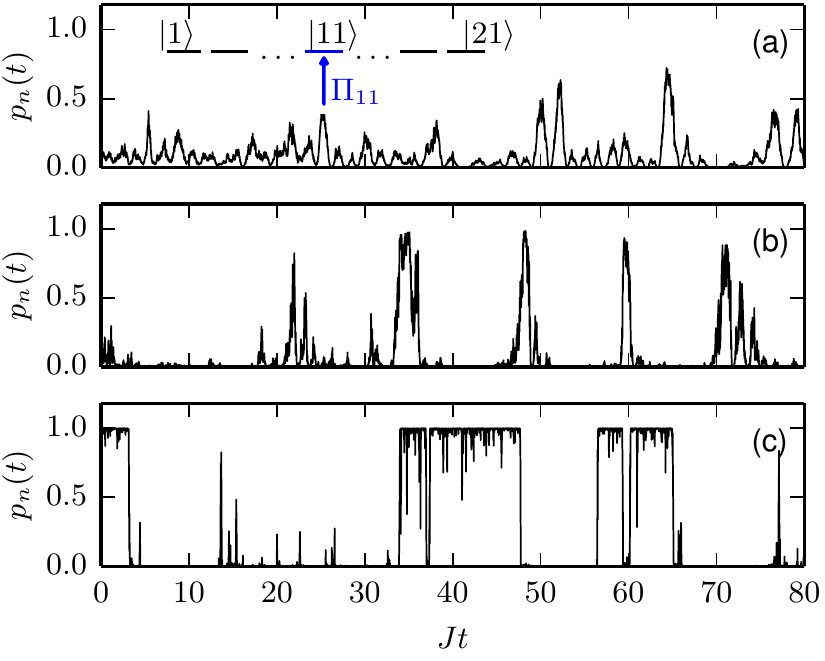}
  \caption{Time evolution of the probability $p_n(t)$ for the particle to be on site $n=11$
   for a lattice of size $N=21$, with the particle initially in its ground state. 
The results are obtained by simulations of the probing of the population with strength (a) $k=0.1J$, (b) $k=1J$, (c) $k=10J$. The inset of (a)
 sketches the $N=21$ sites with arrow indicating the probing of the middle site 
 (i.e., a measurement of the observable $\Pi_{11}$).}
  \label{fig:1}
\end{figure}

In simulations with different numbers of sites $N$ we can identify three
qualitatively different dynamical regimes depending on the measurement strength $k$.
These regimes are exemplified in
Fig.~\ref{fig:1}, which shows the time evolution of the population of the probed
site $p_n(t)=\avg{\Pi_n}_\rho$ for $k=0.1J,J,10J$, $N=21$ and $n=11$.
%COMMENT: Choose larger N? What was \Pi_n ?
For weak probing, where $k\ll J$ (Fig.~\ref{fig:1}(a)), we observe a noisy time evolution superimposed on a recurrent oscillatory recovery of population.
For strong probing with $k=J$ (Fig.~\ref{fig:1}(b)), the population shows a more peaked,
oscillatory  dynamics, and for still stronger probing with $k \gg J$ (Fig.~\ref{fig:1}(c)), the population $p_n(t)$ tends to switch randomly between the values $0$ and $1$, except for very sharp scale-invariant
fluctuations (see \cite{Tilloy15} for a recent discussion). The latter regime manifests the so-called Zeno-like behavior, which may be interpreted in terms of measurement back-action \cite{Misra77}, but which can also be explained without accounting for the measurement outcome as a mere effect of the dissipation term \cite{Schaefer,Facchi08}. 
%
%
%%%%%%%%%%%%%%%%%%%%%%%%%%%%%%%%%%%%%%%%%%%%
%%%%%%%%%%%%%%%%%%%%%%%%%%%%%%%%%%%%%%%%%%%%
%
%
\\
Fig.~\ref{fig:1} shows the evolution of the population of the site probed, as inferred from the (simulated) measurement data and the stochastic master equation (7). It is instructive to study the signal associated with the three different regimes shown in Fig.~\ref{fig:1}, and while this time-dependent signal is dominated by noise, we can obtain its power spectrum in frequency domain, defined as,
\begin{equation}
       \label{eq:PowSpec}
       P_n(\omega) = \frac{1}{2\pi T}\Bigl|\int_0^T e^{-i\omega t } \lambda_t[\Pi_n] dt\Bigr |^2,
\end{equation}
where we assume that the signal is accumulated for an interval between $t=0$ and $t=T$.
As we show in the appendix \ref{app}, the correlation
function $\overline{\lambda_t[\Pi_n] \lambda_{t+\tau}[\Pi_n]}$
can be calculated by the
quantum optical theory of photodetection \cite{GardinerZoller}.
According to this theory, two-time noise correlations in the
detected signal are proportional to two-time quantum correlation
functions of the corresponding system operators. The steady state
spectrum reads
\begin{align}
       \label{eq:SS_Spec}
       S_n(\omega) = &\frac{1}{4\pi}\text{Re}\int_{0}^\infty d\tau \  e^{-i\omega \tau }
       \text{tr}[ \Pi_n e^{\mathcal{L}[\Pi_n] \tau } \{\Pi_n, \rho_\text{ss}\} ] \notag \\
%&=  \text{tr}[ \Pi_n e^{\mathcal{L}[\Pi_n] t }\Pi_n ]  \rho^{nn}_\text{ss},%                   =& \frac{1}{\sqrt{2\pi}}\int_{-\infty}^\infty d\tau e^{-i\omega \tau }
%\text{tr}[\Pi_n e^{\mathcal{L}[\Pi_n]t}\Pi_n\ \rho_\text{ss}]  \notag \\
                   =& \frac{1}{4\pi}\text{Re}\ \text{tr} \Bigl[\Pi_n \bigl(\mathcal{L}[\Pi_n]-i\omega\mathbb{1} \bigr)^{-1}\  \{\Pi_n, \rho_\text{ss}\} \Bigr]
           %   g(\tau)  d\tau.
\end{align}
%
%with the stationary correlation function
%$g_n(\tau) = \avg{\Pi_n(\tau)\Pi_n(0)}_\text{ss}$.
Here, the steady state density matrix $\rho_\text{ss}$ is defined as the solution of the ``average''  master equation, $\dot{\rho}_\text{ss}=\mathcal{L}[\Pi_n]\rho_\text{ss}$,
where $\mathcal{L}[O]\rho=(-i/\hbar)[H,\rho]+k\mathcal{D}[O]\rho$ is deterministic
and discards the random measurement outcome, as if $\mu=0$ in Eq.~(7).
Note that the stochastic density matrix, on average, obeys the deterministic Lindblad
master equation and, if the system has been evolving already for some time, the unheralded state is given by $\rho_\text{ss}$, when we 
begin accumulation of data for the calculation of the spectrum. We recall that, while the signal is proportional to the expectation value of the probed system observable, the signal power spectrum (\ref{eq:PowSpec}) 
only represents the population fluctuations, observed in Fig.(1), in a qualitative sense.  

\subsection{Weak measurement regime}

\begin{figure}
  \includegraphics[scale=1]{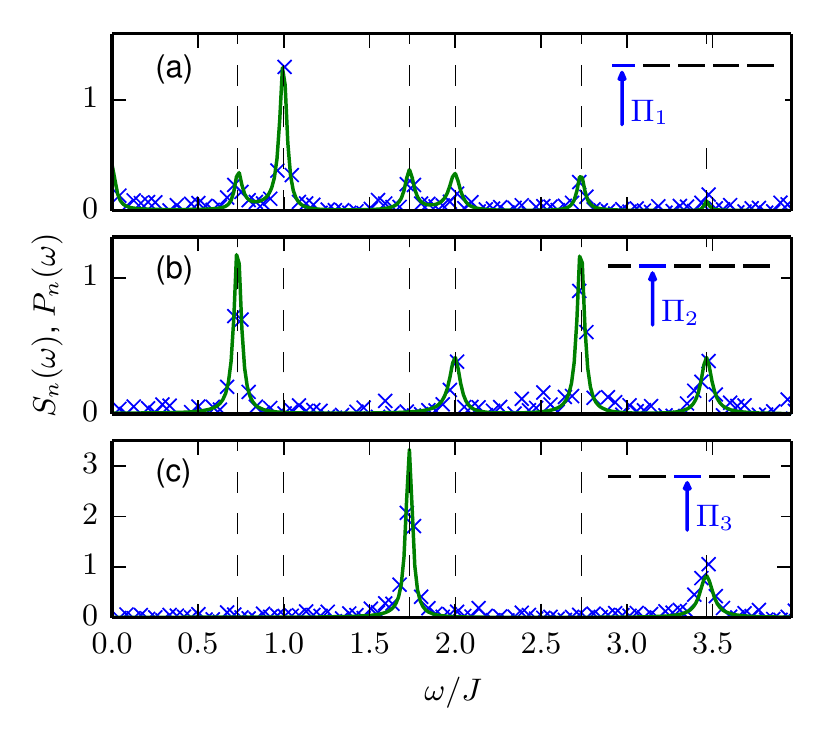}
  \caption{The power spectrum $P_n(\omega)$ for $T=100/J$ averaged over 200 time evolutions
   (blue x) and the steady state spectrum $S_n(\omega)$ (green solid) for $k=0.1J$ for different
    probed sites $n$: (a)~$n=1$, (b)~$n=2$, (c)~$n=3$. The observed frequencies correspond
    to transitions frequencies between different system eigenstates (black dashed).
    The insets sketch the $N=5$ sites with different local probing.}
   \label{fig:2}
\end{figure}

In Fig.~\ref{fig:2}, we show the power spectrum $P_n(\omega)$ averaged over 200
time evolutions with $T = 100/J$ (blue $\times$), and the
calculated steady state spectrum $S_n(\omega)$ (green solid line) for $k=0.1J$ and
three different probed sites ($n=1,n=2,n=3$).
In order to keep the analysis simple, but without loss of generality, we consider $N=5$ sites.
The probing continuously quenches the system and hence induces transient oscillations of the
system observables and hence of the fluctuating measurement signal.
\\
The coinciding spectra $P_n(\omega)$ and $S_n(\omega)$ in Fig.~\ref{fig:2}(a)-(c) hence show sharp spectral peaks
centered at the different Bohr frequencies, $\omega_{ij}=(e_i-e_j)/\hbar$ linking the different eigenstates of the system.
%
%
%%%%%%%%%%%%%%%%%%%%%%%%%%%%%%%%%%%%%%%%%%%%%%%%%%%%%%%%%
%%%%%%%%%%%%%%%%%%%%%%%%%%%%%%%%%%%%%%%%%%%%%%%%%%%%%%%%%
%
%
The extent to which the different transition frequencies appear in the probe signal on a given site depends strongly on which site is measured, and in Fig.~\ref{fig:2}(b) and
(c) some frequencies do not appear at all.
A closer inspection reveals that for the absent frequency components $\omega_{ij}$,
the corresponding amplitude $\bra{w_i}\Pi_n \ket{w_j}$ vanishes, i.e.,
at least one of the eigenstates involved has a node
on the probed site and is not probed and hence not excited
by the measurement back-action.
\\ 
In the absence of probing, the
Liouvillian in Eq.~\eqref{eq:SS_Spec} is
diagonal in the operator basis of dyadic products $\ket{w_j}\bra{w_i}$ (see also the discussion accompanying Fig.(\ref{fig:8})). For weak probing, 
perturbation theory leads to the modified eigenvalues
$\lambda_{ij} = i(e_i-e_j) -\Gamma_{ij}$, with the perturbative
rates $\Gamma_{ij}=k\text{tr}[\ket{w_j}\bra{w_i}\mathcal{D}[\Pi_n](\ket{w_i}\bra{w_j})]$ (non-degenerate case).
Up to a pre-factor, the steady state spectrum then becomes
\begin{equation}
  \label{eq:anal_spec}
   S_n(\omega) \propto \sum_{ij} \frac{\Gamma_{ij}\bra{w_j}\Pi_n\ket{w_i}\rho_\text{ss}^{nn} }{(\omega_{ij}-\omega)^2+\Gamma_{ij}^2},
\end{equation}
where $\rho_\text{ss}^{nn}=\bra{n}\rho_\text{ss}\ket{n}$.
Note that some transitions have the same
energy separation causing  the different peak heights in Fig.~\ref{fig:2}.
\\
Interestingly, probing of the middle site occupation effective acts as a quantum non-demolition measurement of the state parity, 
because neither the Hamiltonian nor the probing couples the odd and the even states.
This has the further consequence, that the probed system does not have a unique steady state, 
and averaged over many realizations of the measurement, the mean occupation on the odd and even subspaces will at any time be given by their initial values.
Moreover, the odd subspace component evolves in a purely unitary manner 
from the initial condition, because its population at the 
probed middle site vanishes. 
The measurements can in this case be used to probabilistically filter out a subset of state components and in some cases to herald the system in special pure superposition states.

\begin{figure}
  \includegraphics[scale=1]{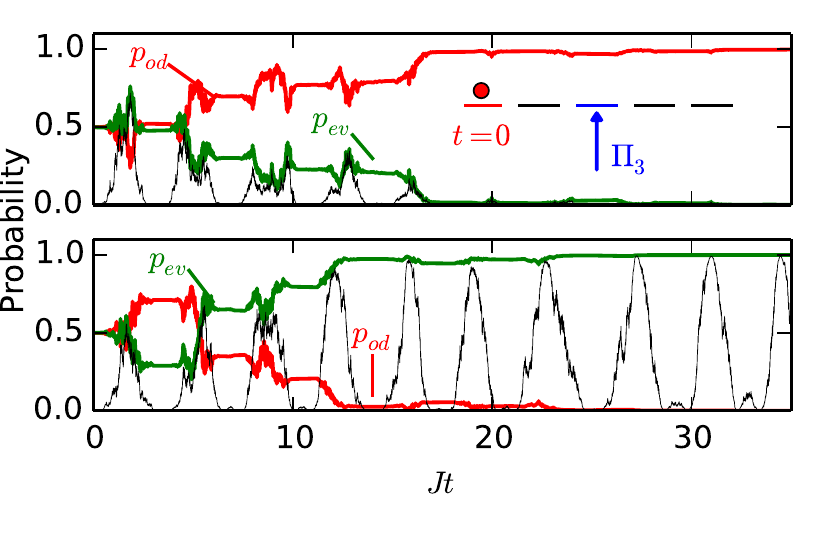}
  \caption{Time evolutions for initial state $\ket{1}\bra{1}$, $N=5$ sites
   and probing on site $n=3$. The black thin line shows the probability
   for the particle the be on the probed site $p_n(t)$, while the red and green
    lines show the probabilities $p_{od}$ and $p_{ev}$ 
    to be in the odd and even subspaces, respectively,
    Despite identical initial conditions, in the upper plot the random measurement
    back-action drives the system into the odd subspace,
    while in the lower case the system converges into the even subspace. }
   \label{fig:3}
\end{figure}

In Fig.~\ref{fig:3}, we show examples of the time evolution when the middle site is measured, starting from a particle localized at the first site, $\rho(0)=\ket{1}\bra{1}$, i.e., in a state with equal weights on the odd and the
even subspace. Repeating the propagation several times, half of the time evolutions end up in the
odd subspace, witnessed by the disappearance of the temporal modulations of the population on the middle site (Fig.~\ref{fig:3}(a)), while half end up in the even
subspace  (Fig.~\ref{fig:3}(b)). In the latter case one observes an oscillatory population on the middle site with the same frequency components as found in the measurement signal, shown in Fig.~\ref{fig:2}(c).

%
%
%%%%%%%%%%%%%%%%%%%%%%%%%%%%%%%%%%%%%%%%%%%%%%%%%%%%%%%%%
%%%%%%%%%%%%%%%%%%%%%%%%%%%%%%%%%%%%%%%%%%%%%%%%%%%%%%%%%
%
%
%%%%%%%%%%%%%%%%%%%%%%%%%%%%%%%%%%%%%%%%%%%%%%%%%%%%%%%%%
\begin{figure}
  \includegraphics[scale=1]{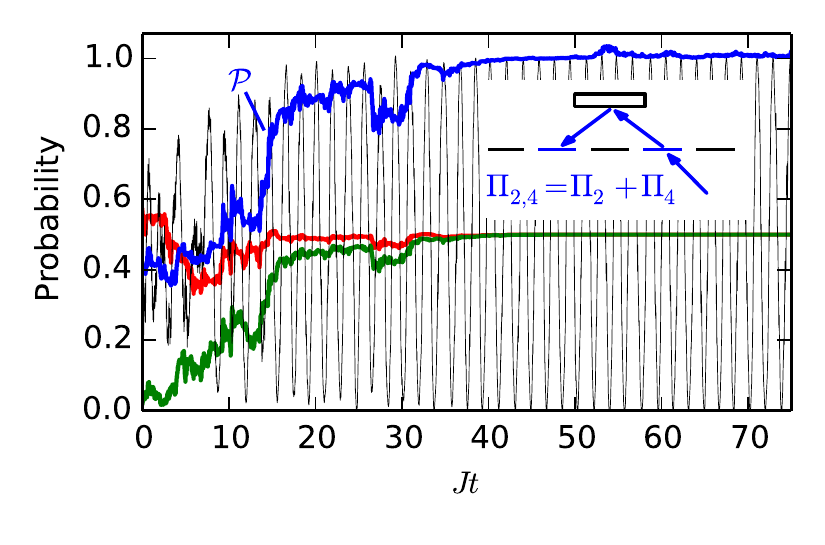}
  \caption{Time evolution for an initial thermally mixed state $\rho(0)=1/Z \exp(-\beta H)$, $\beta=1/J$,
           on $N=5$ sites, where the total population $\Pi_{2,4}=\Pi_2+\Pi_4$ of sites $n=2$ and $n=4$ is probed. Probing of two sites populations can, for example, be realized 
           using a reflecting mirror which sends the probe beam twice through the system 
           (see inset).
           The red and green lines
           show the occupation of the lowest and highest energy eigenstates
           $\bra{w_1}\rho\ket{w_1}$ and $\bra{w_5}\rho\ket{w_5}$, both converging to $0.5$ in the simulation.
           The blue line depicts the purity of the state $\mathcal{P}=\text{tr}[\rho^2]$; its convergence to unity shows that the 
           continuous probing drives the system to a coherent superposition of
           $\ket{w_1}$ and $\ket{w_5}$ in accordance with the full contrast  sinusoidal oscillations
           of the expectation value $\avg{\Pi_{2,4}}$ with
           frequency $(e_5-e_1)/\hbar$ (black line).}
   \label{fig:3a}
\end{figure}
%%%%%%%%%%%%%%%%%%%%%%%%%%%%%%%%%%%%%%%%%%%%%%%%%%%%%%%%%
%

An intriguing dynamics is
obtained by optically interrogating more than one site 
such that their total population is probed.
Consider, for example the situation depicted in the inset of Fig.~\ref{fig:3a}, where two-site
total occupation $\Pi_{2,4}=\Pi_2+\Pi_4$
is measured on a lattice with $N=5$. 
In this case, all simulations eventually lead to a pure conditioned state, being either $\ket{w_3}$, an equal population superposition of the even states $\ket{w_1}, \ket{w_5}$ or an equal population superposition of 
the odd states
$\ket{w_2}, \ket{w_4}$. These three situations occur with probabilities equal to the initial populations of the corresponding subspaces.
As in the previous example, the convergence to the state $\ket{w_3}$, which has nodes on the sites probed, results from 
the detection of a signal with no periodic modulation. 
The distinction between the other two alternatives occurs, because one or the other frequency component $(e_5-e_1)/\hbar$  or $(e_4-e_2)/\hbar$
randomly dominates the measurement signal. 
By measurement back action, the population of the corresponding pair of states, increases and 
shows an oscillatory occupation, as they interfere constructively or destructively on the sites  $n=2$ and $n=4$. 
The random phase of the observed oscillation governs the relative phase on the two states in the superpostion state.
   
In Fig.~\ref{fig:3a}, we consider the chain with $N=5$ sites,
initialized in a thermal state $\rho(t=0)=1/Z \exp(-\beta H)$, with the
partition function $Z=\text{tr}[\exp(-\beta H)]$ and the
inverse temperature $\beta=1/J$, and we show the outcome of a single simulation of the continuous probing of 
the two-site observable
$\Pi_{2,4}$.
The figure shows the stochastic dynamics of the observable $\avg{\Pi_{2,4}}$
and the probability to be in the eigenstates
$\ket{w_1}$ and $\ket{w_5}$ as well as the purity of the state.
\\
If more than one site is probed with the same
light beam a candidate expression for the average measured spectrum, Eq.~\eqref{eq:anal_spec}
reads
\begin{equation}
  \label{eq:anal_spec2}
   S_\mathcal{M}(\omega) \propto \sum_{ij}\sum_{n,m\in\mathcal{M}}
   \frac{\Gamma_{ij}\langle w_j\mid n\rangle \langle m \mid w_i\rangle
   \rho_\text{ss}^{nm} }{(\omega_{ij}-\omega)^2+\Gamma_{ij}^2},
\end{equation}
where $\mathcal{M}$ are the measured sites. We should remember, however, that the observed system does not reach a steady state, 
independent of its initial state and the measurement record, and while our example may suggest
two spectral peaks at frequencies $|e_5-e_1|$ and $|e_4-e_2|$, in
a single realisation, only one of these frequencies
is observed. Repeating the process several times
will sample the different peaks with probabilities reflecting the initial occupation of the corresponding
states and, e.g., allow us to determine the initial temperature.

\subsection{Strong measurement regime}

Increasing the probing strength to $k=J$, so that it becomes comparable
to the systems internal coupling, the trajectories
of the probed site change qualitatively from an oscillatory dynamics
to a quasi-periodic sequence of peaks
(cf.~Fig.~\ref{fig:1}(a),(b)).
%
%
%%%%%%%%%%%%%%%%%%%%%%%%%%%%%%%%%%%%%%%%%%%%%%%%%%%%%%%%%
\begin{figure}
  \includegraphics[scale=1]{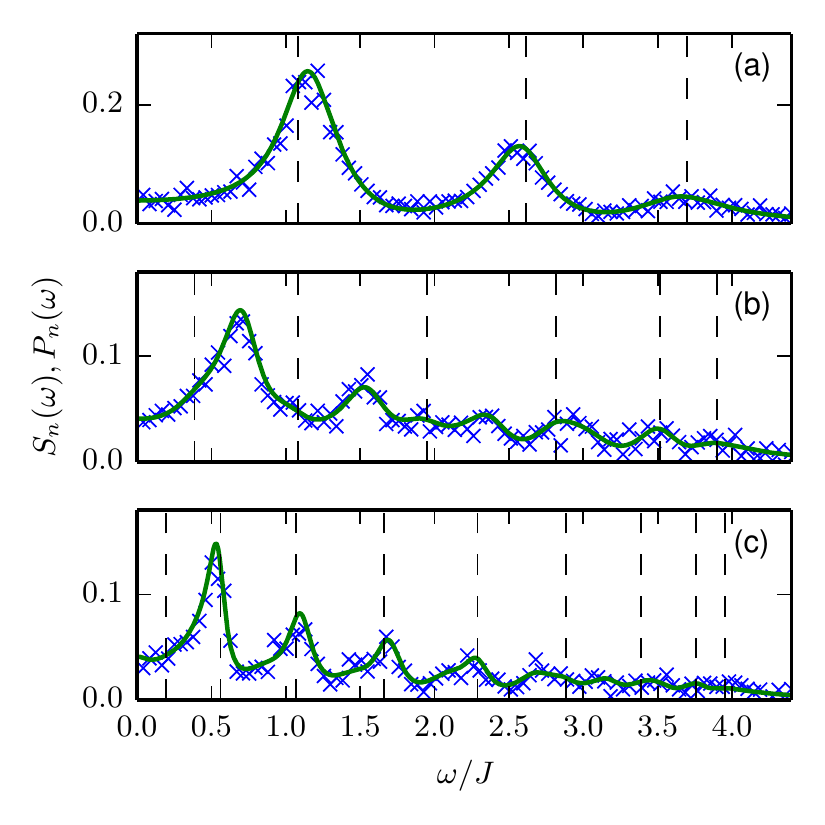}
  \caption{The power spectrum $P_n(\omega)$ for $T=100/J$ averaged over 200 time evolutions
   (blue x) and the steady state spectrum $S_n(\omega)$ (green solid) for $k=J$ and different
    numbers of sites $N$: (a)~$N=7$, (b)~$N=13$, (c)~$N=19$, where the middle site
    $n=N/2+1$ is probed. In all plots a dominant low frequency peak is emerging.
    Moreover the peaks are shifted away from their expected position at the
    transition frequencies between (even) eigenstates (black dashed) towards higher
    frequencies.}
   \label{fig:4}
\end{figure}
%%%%%%%%%%%%%%%%%%%%%%%%%%%%%%%%%%%%%%%%%%%%%%%%%%%%%%%%%
%
%
%
%
%%%%%%%%%%%%%%%%%%%%%%%%%%%%%%%%%%%%%%%%%%%%%%%%%%%%%%%%%
\begin{figure}
  \includegraphics[scale=1]{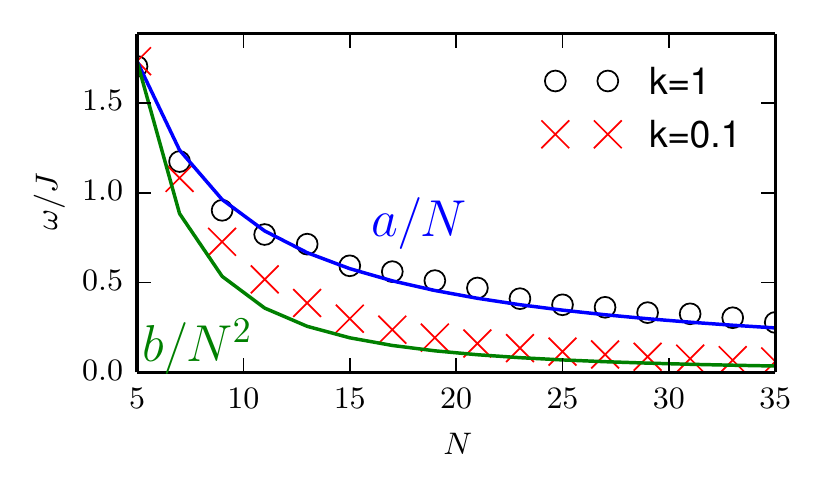}
  \caption{Lowest frequency spectral peak in the measurement signal as function of the number of lattice sites $N$. For $k=0.1$, this frequency decreases
            like $1/N^2$, reflecting the lowest Bohr excitation frequency between the eigenstates. For $k=1$, it scales like $1/N$ as expected for a classical particle bouncing back and forth in a lattice of length $N$. The free parameters were chosen as $a\approx 8.66$ and $b \approx 43.3$.}
   \label{fig:5}
\end{figure}
%%%%%%%%%%%%%%%%%%%%%%%%%%%%%%%%%%%%%%%%%%%%%%%%%%%%%%%%%
%
%
%
%%%%%%%%%%%%%%%%%%%%%%%%%%%%%%%%%%%%%%%%%%%%%%%%%%%%%%%%%
\begin{figure}
  %\hspace{-2.1cm}
  \includegraphics[scale=1]{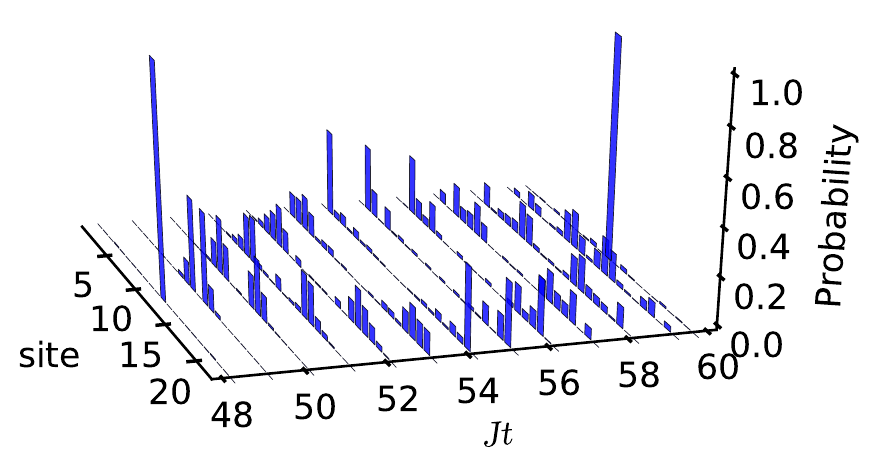}
  \caption{Time evolution between two subsequent peaks
           in Fig.~\ref{fig:1}(b). The measurements localize the
           particle on the measured site and thus creates wave packets
           propagating along the chain. After reflection the wave packets cross
           the probed site again where the measurement back-action refocusses the
           dispersed wave packets.}
   \label{fig:6}
\end{figure}
%%%%%%%%%%%%%%%%%%%%%%%%%%%%%%%%%%%%%%%%%%%%%%%%%%%%%%%%%
%
In Fig.~\ref{fig:4}, we show the power spectrum $P_n(\omega)$ of the simulated signals (the middle site is probed) and the calculated steady state spectrum $S_n(\omega)$ for $k=J$ for different numbers of sites.
In contrast to the case for weak probing (cf.~Fig~\ref{fig:2}),
the amplitudes of the spectral peaks decrease with frequency.
Moreover, with an increasing number of sites, the peaks shift away from
their position at the transition frequencies between eigenstates (vertical dashed lines)
towards higher frequencies.
To understand this result we extract the frequency of the dominant peak,
i.e., the one with the lowest frequency,
as a function of the total number of sites $N$ and compare with
the corresponding lowest frequency peak for weak probing with $k=0.1$ (see Fig.~\ref{fig:5}).
For weak probing, the frequency decreases as $1/N^2$ for large $N$, reflecting the lowest (first) energy gap of the spectrum  in Eq.~\eqref{eq:eval}.
For strong probing, however, it decreases like $1/N$ which can be associated
with the ballistic propagation of a classical particle or a localized wave packets
along the chain. 
Figure \ref{fig:6} shows the population dynamics over the whole chain between two consecutive peaks in the temporal population of the site $n=10$ in 
Fig.~\ref{fig:1}(b).
Indeed, after a measurement has induced a localized peak in the site distribution, one observes
two wave packets moving towards the ends of the chain.
This resembles recent experimental results
on quantum walks observed with cold atoms \cite{Preiss15}.
In our case, however, no specific preparation of the initial
state is needed since the continuous probing
stochastically localizes the particle at the measured site
at some time.
When the wave packet reaches the end of the chain it gets reflected and
travels back to the middle where the probing leads to a
refocusing of the packet and the evolution continues.
During their evolution, the wave packets spread and develop
side peaks explaining the  broad background and higher order
resonances in the spectra (cf. Fig.~\ref{fig:4}).
In this analysis, we considered the case where the middle site was measured.
Probing of a random site will lead to further nontrivial interference
effects between wave packets arriving at different times from different sides of the chain
resulting in more complicated spectra.

\subsection{Zeno regime}

Increasing the measurement strength further, the particle gets effectively projected and stays
on or off the measured site for finite intervals time (cf.~Fig~\ref{fig:1}(c)).
This behavior is associated with the quantum Zeno effect, where the
population transfer between discrete states is inhibited due
to strong or frequent measurement \cite{Misra77}.

\begin{figure}
  \includegraphics[scale=.9]{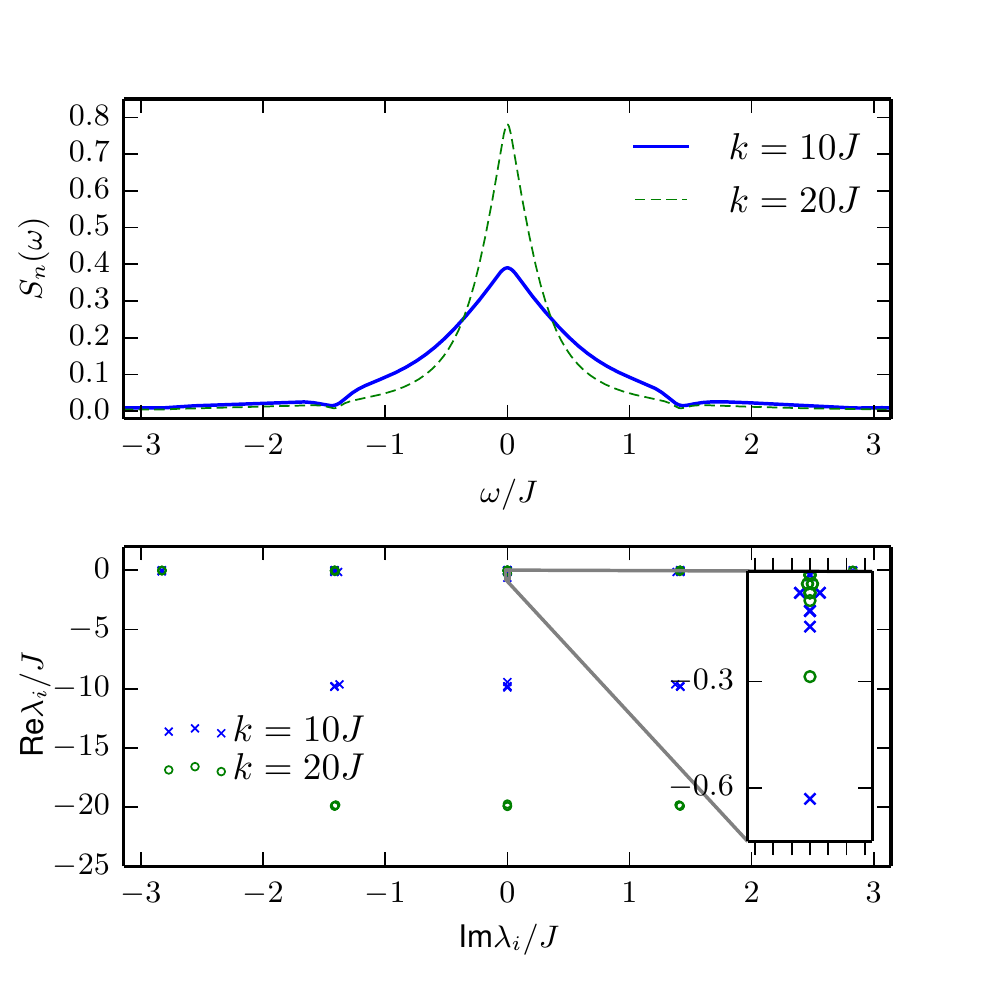}
  \caption{Top: Steady state spectrum in the Zeno regime with $N=9$ and $n=5$.
           Bottom: Eigenvalues of the Liouvillian operator $\mathcal{L}[\Pi_n]$
                    in the complex plan. The eigenvalues appear in groups 
                    with $\text{Re} \lambda_i=-k$ and $\text{Re} \lambda_i=0$. } 
   \label{fig:7}
\end{figure}

Formally, this behaviour stems from
the fact that for short times, the survival probability $p_0$ on an initially populated state varies quadratically in time \cite{Peres80}.
For a closed system, prepared in an eigenstate $\ket{\Psi_0}$ of the measurement operator, we have
\begin{align}
  \label{eq:survive}
  p_0(\delta t)  &= |\bra{\Psi_0} e^{- i H \delta t} \ket{\Psi_0}|^2 \notag \\
              & \approx 1 - \frac{\delta t^2}{\tau_0^2},
\end{align}
where we introduced the time scale $\tau^{-1}_0=\avg{H}_0^2 - \avg{H^2}_0 $,
with $\avg{\dots}_0=\bra{\Psi_0}\dots \ket{\Psi_0}$.
Performing a single measurement $\Pi_n$ after the time $\delta t$ projects
the state back on $\ket{\Psi_0}$ with probability $p_0$
so that after $M$ measurements at intervals $\delta t =t/M$
we have \cite{Facchi08}
\begin{align}
  p^{(M)}_0&(t)  = (p_0(\delta t))^M = \Bigl(1 - \frac{ t^2}{(M\tau_0)^2}\Bigr)^M \notag \\
              & \overset{M \text{large}}{=} \exp(- t^2/M \tau^2_0) 
               = \exp(-t \delta t/\tau_0^2).
\end{align}
We observe the Zeno frozen dynamics ($p^{(M)}_0(t) \rightarrow 1$)  when the probing intervals $\delta t\rightarrow 0$.
Continuous probing with strength $k$ is equivalent \cite{Schulman98} to such repeated 
projective measurement intervals
\begin{equation}
  \delta t \sim 4/k,
\end{equation}
and we can describe the escape of population from an initially localized state by
the effective law %\cite{Pascarizo14}
\begin{equation}
   \label{eq:eff_dec}
   p_n^\text{decay}(t) = \exp(-\gamma_\text{eff}t)
\end{equation}
with $\gamma_\text{eff} = 4/(\tau_0 k)=  4 J^2/k$.
\\
%In order to analyse the dynamical properties of
%the 1D chain in the Zeno regime it is illuminating
%to consider the steady state correlation function
%%
%%
%\begin{align}
%   \label{eq:corr}
%   g(\tau) &= \text{tr}[ \Pi_n e^{\mathcal{L}[\Pi_n] t }\Pi_n \rho_\text{ss} ] \notag \\
%           &=  \text{tr}[ \Pi_n e^{\mathcal{L}[\Pi_n] t }\Pi_n ]  \rho^{nn}_\text{ss},
%\end{align}
%%
%which provides the survival probability \eqref{eq:survive}
%in an open system weighted by the steady state population $\rho^{nn}_\text{ss}$,
%which encodes the probability for the particle
%to enter the probed site.
%Since the steady state spectrum \eqref{eq:SS_Spec} is defined as the Fourier
%transform of Eq.~\eqref{eq:corr} it is appropriate
%to analyse the spectral properties of the system
%in the Zeno regime.
\\
It is instructive to address the power spectrum of the measurement signal near and in the Zeno regime.
In Fig~\ref{fig:7}(top), we thus show the steady state spectra for different values of $k$.
The approximate Lorentzian peaks centered at $\omega=0$,
reflect that the two-time correlation function of the atomic population of the occupied site falls off in an exponential manner, i.e., as if the system prepared on the site $n$ at time $t$ has an exponentially decaying probability to remain on the site until $t+\tau$, cf., Eq.~\eqref{eq:eff_dec}.
For $k=10J$ the Lorentzian is modified by minor dips, but they vanish for larger $k$.

These features follow from an analysis of the eigenspectrum of the Liouvillian $\mathcal{L}[\Pi_n]$
for $k\gg J$. In Fig.~\ref{fig:7}(bottom), we plot the eigevalues
for $k=10J,20J$ in the complex plane.
As one can see, the spectra are divided in groups of eigenvalues with 
$\text{Re} \lambda_i\approx 0$ and   $\text{Re} \lambda_i\approx -k$.
This follows from the fact that in the Zeno regime the coherent part is a small perturbation to the Lindblad term $\mathcal{D}[\Pi_n]$, dominating the
Liouvillian.
The latter is diagonal in the site basis \{$\ket{m}\bra{l}$\} and has the two real
degenerate eigenvalues, $\lambda_i=-k$, for $m\neq n = l$ (or  $l=m\neq n$)
and $\lambda_i=0$ otherwise. It follows that in the first group, there
are $2N-1$ eigenvalues and in the latter one $(N-1)^2$. Moreover,
the zero eigenvalue (with both real and imaginary part vanishing approximately)
is $N-$fold degenerate.
Note that, because the $\lambda_i=-k$ subspace is composed of off-diagonal
operators, its contribution in Eq.~\ref{eq:SS_Spec}
vanishes so that is has no contribution to the detected signal.

\begin{figure}
  \includegraphics[scale=.9]{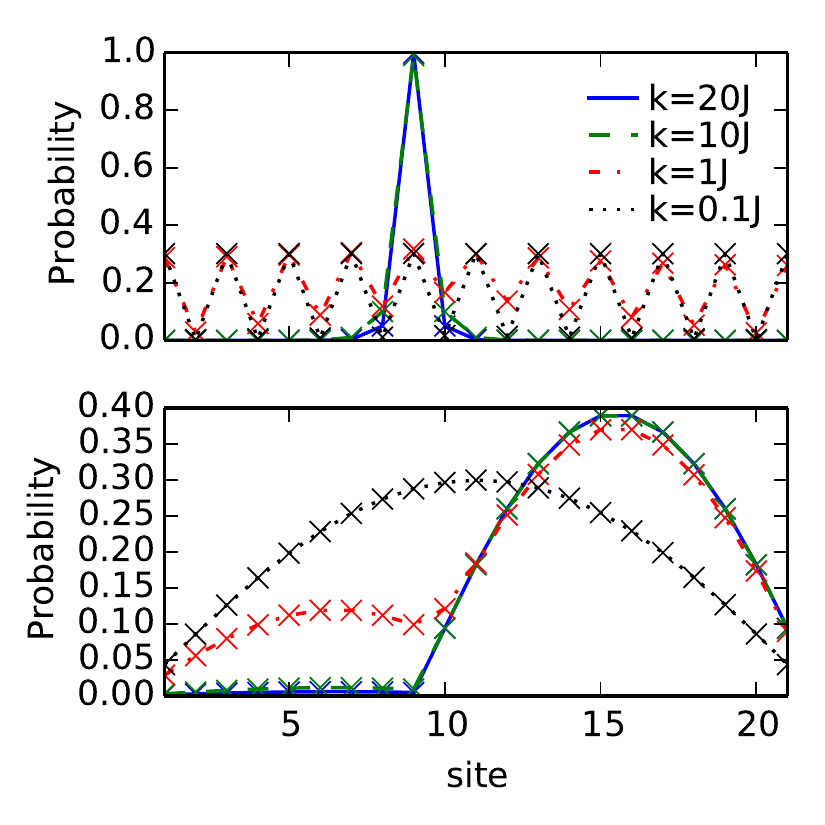}
  \caption{Two effective eigenfunctions $|\langle i\mid \tilde{w}_l\rangle|^2$ defined via
           Eq.~\eqref{eq:eff_eig} for $N=21$, $n=8$ and different measurement
           strength $k$. For $k\rightarrow 0$, the eigenfunction in the lower
           plot coincides with the ground state, while the eigenfunction in the upper
           corresponds to a highly excited energy.
           When $k$ is increased these eigenstates get distorted:
           the eigenfunction in the upper plot localizes 
           on the probed site, while the one in the lower plot 
           develops a node on that site.}
   \label{fig:8}
\end{figure}

In order to visualize the transition from the coherent oscillation regime to the 
Zeno regime, when $k$ increases to values much larger than $J$, 
we note that the deterministic part of Eq.~\eqref{eq:SME}
can be written 
\begin{equation}
\label{eq:nonh}
 d\rho = -\frac{i}{\hbar} \bigl(\tilde{H}\rho - \rho\tilde{H}^{\dagger} \bigr) 
         + 2k \Pi_n \rho \Pi_n,
\end{equation}
where $\tilde{H}=H-ik\Pi_n.$ The eigenstates of the first term in \eqref{eq:nonh}
are dyadic products $\ket{\tilde{w}_i}\bra{\tilde{w}_j}$, where $\ket{\tilde{w}_i}$
are the eigenstates of the effective non-Hermitian Hamiltonian,
\begin{equation}
    \label{eq:eff_eig}
   (H - i k \Pi_n) \ket{\tilde{w}_l}= \tilde{\lambda_l}   \ket{\tilde{w}_l}.
\end{equation}
%
%
%%%%%%%%%%%%%%%%%%%%%%%%%%%%%%%%%%%%%%%%%%%%%%%%%%%%%%%%%
%%%%%%%%%%%%%%%%%%%%%%%%%%%%%%%%%%%%%%%%%%%%%%%%%%%%%%%%%
%
In fact, $\rho_i=\ket{\tilde{w}_l}\bra{\tilde{w}_m}$ offer excellent approximations of the eigenstates of the Liouvillian.
While for small $k$  $\bra{\tilde{w}_i}$ coincide with
the eigenstates of $H$ Eq.~\eqref{eq:evec}, they get considerably deformed
for $k=J$, as shown in Fig~\ref{fig:8} for two specific eigenstates. 
In the Zeno regime ($k\gg J$), the states $\ket{\tilde{w}_l}$
are divided into two orthogonal subspaces,
one containing a single localized state and one containing the remaining states which
all develop a node at the measured site.
\\
It follows, that the main contribution to
the two-time correlation function Eq.~\eqref{eq:SS_Spec} and therefore to the
steady steady spectrum stems from the eigenvalue $\lambda_i$ of the
localized state $\ket{\tilde{w}_i}\bra{\tilde{w}_i}\approx \ket{n}\bra{n}$
because all other states are suppressed in the spectral decomposition
of $\mathcal{L}[\Pi_n]$. The coherent part of $\mathcal{L}[\Pi_n]$
couples the $\text{Re}\lambda_i=0$ and $\text{Re}\lambda_i=-k$ states and perturbs their eigenvalues.
Estimating this shift using 2nd order perturbation theory
suggests $\lambda_i \sim  J^2/k$ which, indeed, 
confirms the scaling of the Zeno rate $\gamma_\text{eff}$ with the system parameters.

%
%
%%%%%%%%%%%%%%%%%%%%%%%%%%%%%%%%%%%%%%%%%%%%%%%%%%%%%%%%%
%%%%%%%%%%%%%%%%%%%%%%%%%%%%%%%%%%%%%%%%%%%%%%%%%%%%%%%%%
%

\section{Conclusion and Outlook}

\label{sec:Summary}
In this article, we have studied the dynamics of
a single quantum particle hopping on a one dimensional lattice and how it is modified by local continuous measurement.
We simulated the conditioned dynamics
using a stochastic master equation, and we analyzed the resulting quantum trajectories in terms of the measurement record and the time-dependent occupation of the probed lattice site. 
\\
We identified three different dynamical regimes:
While for weak measurement strength $k$ the local dynamics
is characterised by almost coherent oscillations
associated with the stochastic preparation of the system
in a superposition of eigenstates, one observes
quasi-periodic oscillation when the measurement strength
reaches the systems energy scale.
The latter behavior can be related to the
ballistic spreading and measurement induced refocussing of the
single particle wave packet.
If the measurement strength is much larger
than the systems internal energy scale, a Zeno-type
dynamics emerges, where the particle localizes on, or off the measured site for
finite time intervals.
\\
While our simple model system displays a variety of phenomena
due to the interplay of the continuous probing and the coherent spatial evolution on the lattice, we expect even more fascinating effects to emerge with
more complicated systems and settings. 
For example, we envision that probing of
motion in 2D lattice systems with complex tunneling amplitudes equivalent to magnetic flux terms, may allow
studies of the robustness and dynamics of the quantum Hall effect or topological states \cite{Hofstadter,Jaksch,Dalibard} under measurements.

There is a growing interest in extending studies of measurement dynamics to spatially extended,
multi-level and many-body systems. Recently, there has thus been a growing interest in the question
to what extent measurement back-action might be a useful tool to engineer states \cite{Hauke13,Stannigel14,Elliott15,Wade15} and to study dynamical features \cite{Lee14,Mazzucchi16} and influence phase transition dynamics \cite{Gammelmark10,Rogers14,Lee,CaballeroBenitez15,Mazzucchi15} in many-body systems. 
These more complex systems also hold the potential for quantum control \cite{Erez} as well as
for highly sensitive quantum metrology, for which measurements and measurement back action play a crucial role \cite{Guta,Gammelmark13b}.
While our present analysis deals with only single particle dynamics, and with the localization dynamics of a single spin excitation in a many-body system, we imagine that, e.g., the identification of transitions between coherent and incoherent spatial propagation will be useful for the understanding of quasi-particle propagation and of localization dynamics in probed many-body systems.

%

%--------------------------------------------------------------------
\begin{acknowledgments}
  The authors acknowledge financial support from the Villum foundation.
\end{acknowledgments}

\appendix
\allowdisplaybreaks
\section{Derivation of the steady state spectrum}
\label{app}
In order to prove that the spectra in Eq.~\eqref{eq:PowSpec} and Eq.~\eqref{eq:SS_Spec}
coincide in the steady state, we follow the lines of \cite{WisemanMilburn} and
calculate the average signal correlation function $\overline{\lambda_t\lambda_{t+\tau}}$ explicitly.
Assuming that $\rho(t)=\rho_\text{ss}$ the detection of a random $\lambda_t[O]$ at
time $t$
leads to the conditioned state
\begin{align}
  \rho_{|\lambda_t} =& \bigl( 1 - \sqrt{2k\mu} \avg{O+O^\dagger}_\text{ss} dW_t \bigr) \rho_\text{ss} \notag\\
                     &+\sqrt{2k\mu} \bigl\{ O,\rho_\text{ss} \bigr\} dW_t.
\end{align}
Evolving this state until $t+\tau$  yields the averaged
density matrix
\begin{align}
  \overline{\rho_{|\lambda_t}}(t+\tau) =& e^{\mathcal{L}[O]\tau}    \rho_{|\lambda_t} \notag \\
                            = & \bigl( 1 - \sqrt{2k\mu} \avg{O+O^\dagger}_\text{ss} dW_t \bigr)
                               \rho_\text{ss} \notag\\
                             &+\sqrt{2k\mu}\ e^{\mathcal{L}[O]\tau} \{O,\rho_\text{ss}\}
                              dW_t.
\end{align}
At time $t+\tau$ the measurement outcome is then given
\begin{align}
\lambda_{t+\tau}[O] =& \avg{O}_\text{ss}
                   \bigl( 1- \sqrt{2k\mu}\avg{O+O^{\dagger}}_\text{ss}dW_t\bigr)dt \notag \\
                  & +\sqrt{2k\mu}\ \text{tr} \bigl[O\ e^{\mathcal{L}[O]\tau}
                  \{O,\rho_\text{ss} \} \bigr] dW_t dt                  \notag  \\
                  &+ \frac{1}{\sqrt{8 k}} dW_{t+\tau}.
\end{align}
Finally, we multiply $\lambda_t[O]$ with $\lambda_{t+\tau}[O]$ and average over
different realizations of $dW_t$ leading to the correlation function
\begin{align}
\overline{\lambda_t[O] \lambda_{t+\tau}[O]} =& \avg{O}^2_\text{ss} dt^2 - \frac{\sqrt{\mu}}{2}
                                          \avg{O}_\text{ss} \avg{O+O^\dagger}_\text{ss}
                                          dt^2 \notag  \\
                                       & +\frac{\sqrt{\mu}}{2}
                                      \text{tr} \bigl[Oe^{\mathcal{L}[O]\tau} \{O,\rho_\text{ss}\}\bigr]
                                         dt^2,
\end{align}
where we used that $dW_t^2=dt$ and $\overline{dW_t}=\overline{dW_tdW_{t+\tau}}=0$.
The $\tau$-dependent term (with $O=\Pi_n$) is precisely the one entering the expression
\eqref{eq:SS_Spec} for the spectrum.

\bibliography{literature}

\end{document}